\documentclass{article}

\usepackage[english]{babel}
\usepackage[utf8x]{inputenc}
\usepackage[T1]{fontenc}
\usepackage{geometry} % to change the page dimensions

\usepackage{graphicx}
\usepackage{tikz}
\usepackage{tikz-cd} 
\usepackage{amssymb}
\usepackage{amsmath}
\usepackage{amsthm}

\usepackage{extarrows}
\usepackage{chngcntr}
\usepackage{caption}
\usepackage{subcaption}

\newtheorem{theorem}{Theorem}

\theoremstyle{definition}

\theoremstyle{remark}

\newcommand{\R}{\mathbb{R}}

\usepackage{changes} %<- For revision.
\begin{document}

\title{Dynamics of Crime In and Out of Prisons}

\author{Jongo Park\footnote{carlorton625@unist.ac.k}  and  Pilwon Kim\footnote{pwkim@unist.ac.kr}\\
Department of Mathematical Sciences \\
Ulsan National Institute of Science and Technology(UNIST)\\ 
Ulsan Metropolitan City\\
44919,  Republic of Korea}

\maketitle

\begin{abstract}
	The accumulated criminal records shows that serious and minor crimes differ in many measures and are related in a complex way. While some of those who have committed minor crime spontaneously evolve into serious criminals, the transition from minor crime to major crime involves many social factors and have not been fully understood yet. In this work, we present a mathematical model to describe how minor criminals turns in to major criminals inside and outside of prisons. The model is design to implement two social effects which respectively have been conceptualized in popular terms ``broken windows effect'' and ``prison as a crime school.'' Analysis of the system shows how the crime-related parameters such as the arrest rate, the period of imprisonment and the in-prison contact rate affect the criminal distribution at equilibrium. Without proper control of contact between prisoners, the longer imprisonment rather increases occurrence of serious crimes in society. An optimal allocation of the police resources to suppress crimes is also discussed. 
\end{abstract}

\section{Introduction}
Understanding what factors cause a high crime rate is essential to developing effective measures to prevent crime in a community. The accumulated criminal records \cite{national2009understanding,kesteren2000criminal} shows that serious and minor crimes differ in many measures such as occurrence rate, arrest rate and rehabilitation rate. There is also difference between control activity of the police devoted to serious crimes and that devoted to minor crimes \cite{britt1975crime}.  

In recent years, there have been substantial progresses in developing mathematical tools to investigate criminal activity. Mathematical models based on the reaction-diffusion equations have been proposed \cite{short2008statistical, short2010nonlinear, rodriguez2010local} to study dynamics of localized patterns of criminal activity, especially focusing on the re-victimisation phenomena and the hot-spot formation. Some other models have adapted their basis from population biology, such as infectious disease models \cite{campbell1997social,ormerod2001non, mcmillon2014modeling} and predator-prey models \cite{vargo1966note,comissiong2012criminals,nuno2008triangle}. A similar approach was used for modeling organized crimes \cite{comissiong2012life,sooknanan2013catching}, where gang membership is treated as an infection that multiplies through peer contagion.

In this study, we present a mathematical model to describe how minor criminals turns in to major criminals. While some of those who have committed minor crime spontaneously evolve into serious criminals \cite{trove.nla.gov.au/work/9870833}, the transition from minor crime to major crime involves many social factors and have not been fully understood yet. Besides the basic progressive nature of crime, we are interested in finding extra factors that accelerate the transition from minor crimes to major crimes. This paper focuses on criminal transitions occurring inside and outside of prison, which respectively have been conceptualized in popular terms ``broken windows effect'' and ``prison as a crime school'': the broken windows theory states that accumulation of low level offenses in a community, if not adequately controlled, acts as a social pressure that leads to more serious crimes \cite{wilson1982police, harcourt2006broken, cerda2009misdemeanor}. On the contrary, in prisons, staying with many criminal peers in a limited facility makes minor criminals frequently contact with hard-core and skilled criminals and possibly deepens their illegal involvement \cite{damm2016prison, cullen2011prisons,henneguelle2016better}. The peer-effect on crime recidivism is strongly supported by recent empirical research in many countries \cite{bayer2009building,damm2013deal,ouss2011prison}. %check for grouping of ref.

We are especially interested in the influence of over-crowded prison facilities on criminal transition. A mathematical model reflecting  the effect of incarceration on recidivism has been proposed in  \cite{mcmillon2014modeling}.  However, to the authors' knowledge, the in-prison dynamics between minor and major criminals and the effect of the prison capacity on it have never been studied before in a mathematical framework. The behavior of the model is investigated through stability analysis, bifurcation analysis, and numerical simulations. By analyzing the corresponding system of equations, we demonstrate how the crime-related parameters such as the arrest rate, the rehabilitation rate and the capacity of the prison affect the criminal distribution at equilibrium. The results also suggests an optimal allocation of the police resources to minimize occurrence of serious crimes. This may be used to assist policy-makers in the development of effective crime control strategies.

\section{Model}
\subsection*{Basic assumptions}
We first formulate our proposal as a compartmental model, with the population $T>0$ being divided into five disjoint groups $N, M, F, P_M$  and $P_F$. The group $N$ represents non-criminals who have never involved any criminals or have finished serving their prison sentence.  $M$ and $F$  denote individuals who have committed misdemeanor and felony respectively, but have not get arrested yet. Once they are arrested and are sentenced to be imprisoned, they become inmates,  $P_M$ and $P_F$ , respectively. They may return to normal civilian  $N$ after serving their sentence in prison. However, part of them commit a crime again, reverting back to $M$ or $F$. Figure 1 shows the structure of transitions occurring between groups in the basic model. 
\begin{figure}[h!]
	\includegraphics[scale=0.4]{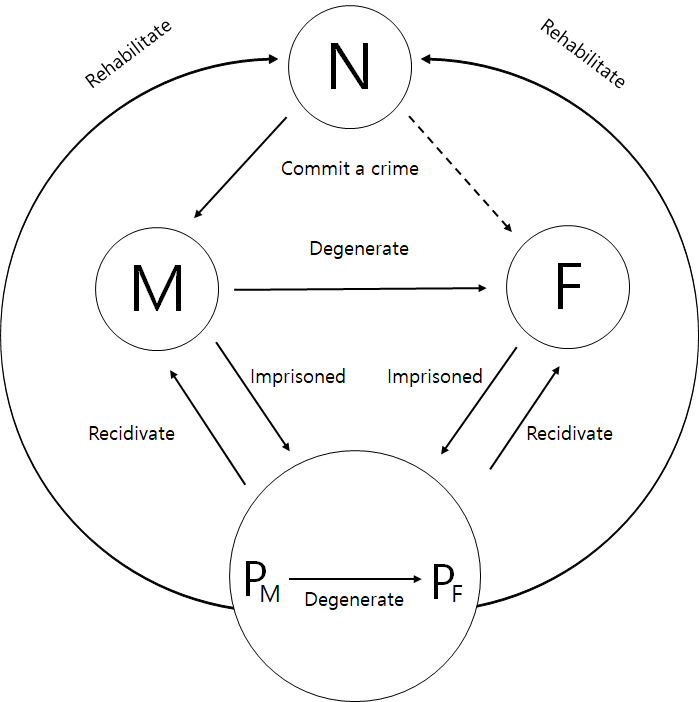}
	\vspace*{8pt}
	\centering
	\caption{Transition diagram for the basis model. The direct transition from $N$ to $F$ is considered negligible and is not implemented in the model.}
%	\caption{{\color{red}Transition diagram for the basis model}}
\end{figure}

In the absence of good evidence to the contrary, we follow the common rules in population dynamics: 1) the transfer out of any particular group is proportional to its size. 2) if the transfer is caused by contact between two group members, it is proportional to the size of both groups. Based on the rule 1, we set $c\ge 0$ and $d\ge 0$ to express the transition rates from $N$ to $M$, and from $M$ to $F$, respectively. 
\begin{equation}
N\xlongrightarrow{   c}M\quad\text{and}\quad M\xlongrightarrow{d}F
\label{N2M2F}
\end{equation}
We assume the direct transition from $N$ to $F$ is negligible compared to that from $M$ to $F$. The sequential transition from $N$ through $M$ to $F$ is justified from the report that most of those people who commit major crime have committed minor crimes before \cite{trove.nla.gov.au/work/9870833}. 

The second rule(transition by contact) is based on the assumption that the population is homogeneously mixed, and should be dealt in our model with care. While the models for organized gang crimes \cite{comissiong2012life,sooknanan2013catching} treat the transition between gang members as contagion by frequent contact, such analogy with epidemic process may be not adequate for daily contacts occurring in a general society. Considering the nature of criminals that hide their true intension from others, we can presume that assimilation with criminals occurring by random contact is rare and negligible compared to other factors that we will consider in the following sections. 

Once criminals are caught and convicted, they are imprisoned to serve their sentence. If the parameters $a_M\ge 0$ and $a_F\ge 0$ are respectively the arrest-and-conviction rates for misdemeanor and felony, the corresponding transitions are 
\begin{equation}
M\xlongrightarrow{a_M}P_M\quad\text{and}\quad F\xlongrightarrow{a_F}P_F
\label{MF2P}
\end{equation}
Let $i_M>0$ and $i_F>0$ be the period of imprisonment for minor and major criminals, respectively. In addition, let $0\le r_M \le 1$ and $0\le r_F \le 1$ denote the rehabilitation rates for minor and major criminals, respectively, which determines the proportion of prisoners moving back to  after release. Then we have the transitions
\begin{equation}
P_M\xlongrightarrow{r_M/i_M}N\quad\text{and}\quad P_F\xlongrightarrow{r_F/i_F}N
\label{P2N}
\end{equation}
This implies that the other portions, $1-r_M$ and $1-r_F$, of inmates commit a crime again, as
\begin{equation}
P_M\xlongrightarrow{(1-r_M)/i_M}M\quad\text{and}\quad P_F\xlongrightarrow{(1-r_F)/i_F}F
\label{P2MF}
\end{equation}

\subsection*{Broken windows effect out of prisons}
Although the above transitions describe the basic structure of dynamics between criminals and non-criminals, it does not properly reflect important interactions between them. We extend the basic model to incorporate the broken windows theory. The theory states that seemingly petty signals of mischief, if not adequately controlled, elicit more serious crime. In order to describe the atmospheric pressure that accelerates transition from minor criminals to major criminals, we add to (1) the quadratic size effect as
\begin{equation}
M\xlongrightarrow{bM}F
\label{MM2FF}
\end{equation}
where $b \ge 0$ is the coefficient that represents the broken-windows effect. 

\subsection*{Crime-school effect in prisons}
We further extend the model to investigate how interactions in prison influence post-release behavior of criminals. The beneficial deterrent effect of prison may be weakened by the negative side-effects of incarceration: Through routine contacts in a limited facility area, criminals get to learn from each other, build new networks and find new opportunities for crime \cite{henneguelle2016better}. 

We assume that, in prison, criminal motivations and skills are spread between inmates and the ``contagion'' occurs by frequent contacts between them. To describe such in-prison transition from the minor criminals to the major criminals, we define $P_M’$ as those who used to be minor criminals and turn to be major criminals by assimilation with them. They emerged from the transitions as 
\begin{equation}
P_M+P_F\xlongrightarrow{\beta}P_M'+P_F\quad\text{and}\quad P_M+P_M'\xlongrightarrow{\beta}P_M'+P_M'
\label{2PMP}
\end{equation}
where $\beta\ge 0$ is the transmission contact rate of prisoners. It is desirable to maintain the total number of the prisoners under the capacity of the prison facilities. Overcrowding in prison increases intensity of interactions between prisoners and raises risk of recidivism \cite{RePEc:hal:journl:halshs-01184046}. The existence of $P_M’$ is problematic, as they are virtually major criminals, while they are treated as minor ones: they are ``disguised major criminals'' and are released after short detention with a low rehabilitation rate.
\begin{equation}
P_M'\xlongrightarrow{r_F/i_M}N\quad\text{and}\quad P_M'\xlongrightarrow{(1-r_F)/i_M}F.
\label{PMP2}
\end{equation}
Now, based on the transitions (1) to (7), we derive the corresponding model as
\begin{equation}
\begin{aligned}
\frac{dN}{dt}&=-cN+\frac{r_M}{i_M}P_M+\frac{r_F}{i_F}P_F+\frac{r_F}{i_M}P_M'\\
\frac{dM}{dt}&=cN-(a_M+d)M-bM^2+\frac{1-r_M}{i_M}P_M\\
\frac{dF}{dt}&=dM+bM^2-a_F F+\frac{1-r_F}{i_F}P_F+\frac{1-r_F}{i_M}P_M'\\
\frac{dP_M}{dt}&=a_M M-\frac{1}{i_M}P_M-\beta P_M(P_F+P_M')\\
\frac{dP_F}{dt}&=a_F F-\frac{1}{i_F}P_F\\
\frac{dP_M'}{dt}&=-\frac{1}{i_M}P_M'+\beta P_M(P_F+P_M')
\label{model}
\end{aligned}
\end{equation}
where $N\left(0\right)\geq0, M\left(0\right)\geq0, F\left(0\right)\geq0, P_M\left(0\right)\geq0, P_F\left(0\right)\geq0$ and $P_M'\left(0\right)\geq0$. In our work, we assume that the transitions between the compartments of the model converge to equilibrium in short time scales and the society maintains with a fixed total population $T=N+M+F+P_M+P_F+P_M'$ in the meantime. The model does not consider birth and death of population: including these factors requires nontrivial extension of work, since there are substantial difference in the death rate between criminals and non-criminals. It has been reported that incarceration seriously reduces life expectancy \cite{wildeman2016incarceration}. Some study has shown that five years behind bars increased the chance of death by 78$\%$ and shortened the expected life span at age 30 by 10 years \cite{patterson2013dose}. To involve these factors in the model, one needs to find a possible mechanism that induces such difference in demographic data. We leave this as a plausible extension of the current model for the future study. 
%The model is completed by the identity $T=N+M+F+P_M+P_F+P_M'$ which states that the population is closed.

\section{Analysis of Model}
The model (\ref{model}) is a 6-dimensional nonlinear systems and is hard to analytically study as is. In the following, we investigate the asymptotic behaviours of the model for the four extreme cases; i) no crime-school effect, ii) no broken-windows effect, iii) low arrest rate and strong punishment, and iv) strong precaution and low rehabilitation.

\begin{theorem}[No crime-school effect] Suppose all the parameters in the system (\ref{model}) are positive except $\beta=0$. There is a unique equilibrium $\left(N^{\star},M^{\star},F^{\star},P_M^{\star},P_F^{\star},P_M'^{\star}\right) $ of the system (\ref{model}) such that 
\begin{align*}
	M^{\star}&=\frac{-p+\sqrt{w}}{2q}, \\
	F^{\star}&=\frac{cT-s_M M^{\star}}{s_F}\\
	P_M^{\star}&=i_M a_M M^{\star},\\
	P_F^{\star}&=i_F a_F F^{\star},\\
	P_M'^{\star}&=0,\\
	N^{\star}&=T-M^{\star}-F^{\star}-P_M^{\star}-P_F^{\star},
\end{align*}
where	 
\begin{align*}
	p&=cd+cda_F i_F+c a_F r_F+d a_F r_F+c a_M a_F i_M r_F+ a_M a_F r_M r_F,\\
	q&=b c +b c a_F i_F+b a_F r_F,\\
	w&=p^{2}+4ca_F r_F qT,\\
	s_M&=c+a_M i_M+a_M r_M,\\
	s_F&=c+a_F i_F+a_F r_F.
\end{align*}
	The equilibrium is locally asymptotically stable either for sufficiently small values of $a_M$ and $a_F$ , or for sufficiently large values of $i_M$  and $i_F$.
	\label{BW}
\end{theorem}
Without the crime-school effect, no minor criminals in the prison degenerates to potential felonies, that is, $P_{M}'^{\star}=0$. The population of minor and major criminals in and out of the prison are directly proportional with the arrest rate and the imprisonment period. One can estimate the population of the uncaught criminals from the number of inmates as
\begin{equation}
M^{\star}=\frac{1}{i_M a_M} P_M^{\star}\quad \text{and}\quad F^{\star}=\frac{1}{i_F a_F} P_F^{\star}.
\label{estimate}
\end{equation}
The next theorem shows that the distribution of criminals become more complex if minor criminals in prison can turn into potential felonies. 
\begin{theorem}[No broken-windows effect]
		Let $b=0$. There is a unique equilibrium $\left(N^{\star},M^{\star},F^{\star},P_{M}^{\star},P_{F}^{\star},P_{M}'^{\star}\right)$  of the system of (\ref{model})
		\begin{align*}
		M^{\star}&=\frac{1}{ui_{M}}\left(ca_{F}i_{M}r_{F}T-P_{M}^{\star}s_{F}\right)\\
		F^{\star}&=\frac{1}{ui_{M}}\left(\left(d+a_{M}\left(1-r_{F}\right)\right)ci_{M}T+P_{M}^{\star}\left(dr_{F}-dr_{M}-s_{M}\right)\right)\\
		P_{M}^{\star}&=\frac{-p+\sqrt{w}}{2q}\\
		P_{F}^{\star}&=a_{F}i_{F}F^{\star}\\
		P_{M}'^{\star}&=-P_{M}^{\star}+a_{M}i_{M}M^{\star}
		\end{align*}
		where 
		\begin{align*}
		p&=cd+cda_{F}i_{F}+cd\beta a_{F}i_{M}i_{F}T+c\beta a_{M}a_{F}i_{M}i_{F}T+ca_{F}r_{F}+da_{F}r_{F}+ca_{M}a_{F}+da_{F}r_{F}\\
		&+ca_{M}a_{F}i_{M}r_{F}+c\beta a_{M}a_{F}i_{M}^{2}r_{F}T+a_{M}a_{F}r_{M}r_{F}-c\beta a_{M}a_{F}i_{M}i_{F}r_{F}T\\
		q&=-cd\beta i_{M}-c\beta a_{F}i_{F}-cd\beta a_{F}i_{M}i_{F}-c\beta a_{M}a_{F}i_{M}i_{F}-d\beta a_{F}i_{F}r_{M}\\
		&-\beta a_{M}a_{F}i_{F}r_{M}-c\beta a_{F}i_{M}r_{F}-d\beta a_{F}i_{M}r_{F}-c\beta a_{M}a_{F}i_{M}^{2}r_{F}+c\beta a_{F}i_{F}r_{F}+d\beta a_{F}i_{F}r_{F}\\
		&+c\beta a_{M}a_{F}i_{M}i_{F}r_{F}-\beta a_{M}a_{F}i_{M}r_{M}r_{F}+\beta a_{M}a_{F}i_{F}r_{M}r_{F}\\
		w&=p^{2}+4ca_{M}a_{F}i_{M}r_{F}qT\\
		u&=cd+ca_{M}+cda_{F}i_{F}+ca_{M}a_{F}i_{F}-ca_{M}r_{F}+ca_{F}r_{F}+da_{F}r_{F}\\
		&+a_{M}a_{F}r_{F}+ca_{M}a_{F}i_{M}r_{F}-ca_{M}a_{F}i_{F}r_{F}\\
		s_{M}&=c-cr_{F}+ca_{M}i_{M}+a_{M}r_{M}-ca_{M}i_{M}r_{F}-a_{M}r_{M}r_{F}\\
		s_{F}&=c-cr_{F}+ca_{F}i_{F}+a_{F}r_{F}-ca_{F}i_{F}r_{F}-a_{F}r_{M}r_{F}
		\end{align*}
		The equilibrium is locally asymptotically stable for sufficiently small $a_{M}$ and $a_{F}$.
\label{CrimSchool}
\end{theorem}

One can confirm that $P_{M}'^{\star}>0$ at equilibrium if $b=0$. Since the potential felonies $P_{M}'$ are formally counted as minor criminals, estimate of the criminal population based on the number minor and major criminals as in (\ref{estimate})  results in an undershoot of minor criminals in society.

In the next two theorems we describe asymptotic behaviour of the model (\ref{model}) with small parameters, based on the geometric perturbation theory \cite{tikhonov1952systems}. Theorem 3 deals with the case of the low arrest rate and the long imprisonment for major criminals. This corresponds to the situation that the police fail to track down criminals properly due to limited budget or lack of effective measures, while they make an example of few arrested criminals by punishing them severely.

\begin{theorem}[Low arrest rate and strong  punishment]
		Suppose $a_{F}=k_{1}\epsilon$ and $\frac{1}{i_{F}}=k_{2}\epsilon$ where $0<\epsilon\ll1$ and $k_{1},k_{2}>0$.  Then, the corresponding degenerate system has a solution $\Gamma^{0}=\left(N^{0},M^{0},F^{0},P_{M}^{0},P_F^{0},P_{M}'^{0}\right)$ where
		\begin{align*}
		&N^{0}\left(t\right)=M^{0}\left(t\right)=P_{M}^{0}\left(t\right)=P_{M}'^{0}\left(t\right)=0,\\
		&P_{F}^{0}=\frac{k_{1}}{k_{1}+k_{2}}T+C\exp\left(-\left(k_{1}+k_{2}\right)Tt\right),\quad \text{and}\quad F^{0}\left(t\right)=T-P_{F}^{0}\left(t\right)
		\end{align*}
		for some $C\in\R$. here exists a locally unique attracting solution $\Gamma^{\epsilon}$ of the system $\left(\ref{model}\right)$ for $\epsilon$ sufficiently small, which tends to $\Gamma^{0}$ for $\epsilon\rightarrow0$.
		\label{SFP}
\end{theorem}

Theorem 3 shows that if the arrest rate for major crimes is extremely low, no matter how strong the punishment is, most population turns into major criminals, either in or out of the prison. The next case is for the low transition rate $c$ and the low rehabilitation rates $r_M$ and $r_F$.
One of the possible scenario for this is stigmatizing persons  who have ever committed crimes and hardly accepting them as a part of society. This also gives a signal of strong precaution to people that they can be expelled from the society even with one minor criminal activity, leading to the low transition rate $c$.

\begin{theorem}[Strong precaution and low rehabilitation]
		Suppose $r_{M}=k_{1}\epsilon$, $r_{F}=k_{2}\epsilon$ and $c=k_{3}\epsilon$ where $0<\epsilon\ll1$ and $k_{1},k_{2},k_{3}>0$.  Then, the corresponding degenerate system has a solution $\Gamma^{0}=\left(N^{0},M^{0},F^{0},P_{M}^{0},P_F^{0},P_{M}'^{0}\right)$ where
		\begin{align*}
		&M^{0}\left(t\right)=P_{M}^{0}\left(t\right)=P_{M}'^{0}\left(t\right)=0,\\
		&N^{0}\left(t\right)=\frac{k_{2}a_{F}}{k_{2}a_{F}+k_{3}\left(1+a_{F}i_{F}\right)}T+C\exp\left(-\left(\frac{a_{F}}{1+a_{F}i_{F}}k_{2}+k_{3}\right)t\right),\\
		&F^{0}\left(t\right)=\frac{1}{1+a_{F}i_{F}}\left(T-N^{0}\left(t\right)\right),\quad\text{and}\quad P_{F}^{0}\left(t\right)=\frac{a_{F}i_{F}}{1+a_{F}i_{F}}\left(T-N^{0}\left(t\right)\right)
		\end{align*}
		for some $C\in\R$. here exists a locally unique attracting solution $\Gamma^{\epsilon}$ of the system $\left(8\right)$ for $\epsilon$ sufficiently small, which tends to $\Gamma^{0}$ for $\epsilon\rightarrow0$.
		\label{SocialStigama}
\end{theorem}
Making criminals' rehabilitation hard with strict separation eventually divides the population into non-criminals and felonies. To have a larger potion of non-criminals, however, we need to hold relatively higher level of the rehabilitation rate than the transition rate, that is, $k_2>k_3$.

\section{Bifurcation Analysis}
In this section, we perform the bifurcation analysis of the proposed model (\ref{model}). For a typical simulation, we set the parameters as
\begin{equation}
\begin{aligned}
&b=0.00001,c=0.00012,d=0.0004\\
&a_{M} =0.1,a_{F} =0.1,r_{M} =0.4,r_{F} =0.2\\
&\beta=0.001,i_M=0.5,i_F=5.
\end{aligned}
\end{equation}
These parameters are calibrated such that the distribution of major/minor criminals and their arrest rates largely agree with crime statistics in several countries \cite{national2009understanding,kesteren2000criminal}. We set the total population 
$T=1,000,000$ through out the analysis. 
\begin{figure}[h!]
	\includegraphics[width=0.7\textwidth]{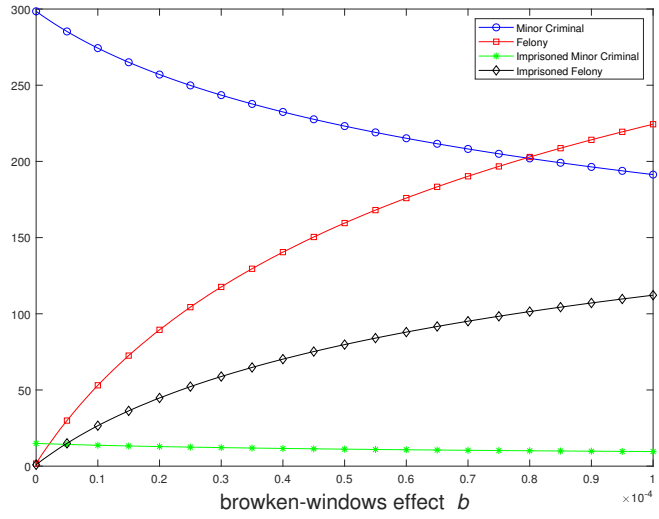}
	\centering
	\vspace*{8pt}
	\caption{Equilibrium distribution according to the broken-windows effect $b$}
\end{figure}
In Figure 2, the bifurcation diagram for the broken-windows effect $b$ is illustrated. The minor crime tends to decrease and the major crime increases as $b$ grows. This implies that elimination of environmental factors that reveal misdemeanors is important to prevent occurrence of more serious crimes. Note that keeping $b$ near zero cuts down major criminals to a negligible level. This implies that the broken widows effect becomes even more important in a safe society where the ratio of major criminals is relatively low.

\begin{figure}[h!]
	\includegraphics[width=0.7\textwidth]{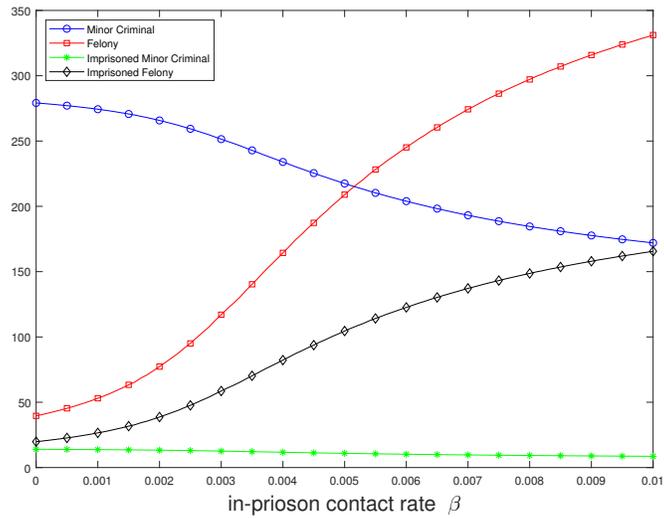}
	\centering
	\vspace*{8pt}
	\caption{Equilibrium distribution according to the in-prison contact rate $\beta$}
\end{figure}

While suppressing $b$ promotes the preventive effect on serious crimes out of prison, controls inside prison can work as a more practical measure against crimes. The bifurcation diagram in Figure 3 shows how the distribution of criminals changes with the contact rate $\beta$ in prison. The rise of $\beta$  increases the number of major criminals in society.

Another in-prison measure to control crimes is the period of imprisonment for criminals. There are mixed evidence regarding the question of whether spending more time in prison increases the rehabilitation rate \cite{reci2, reci3}. However, here we assume that the period of imprisonment and the rehabilitation rate are weakly positively correlated, as long as criminals are held in custody in a effectively managed facility. Let us denote $w \ge 1$ as a general weight of offense and set the period of imprisonment as 
\begin{equation}
i_M=w\, i_M^{\text{min}}\quad \text{and}\quad i_F=w\, i_F^{\text{min}}
\end{equation}
where $i_M^{\text{min}}=0.5$(year) and $i_F^{\text{min}}=5$(year) are the minimum period for minor and major criminals, respectively. The weight of offense is also related with the rehabilitation rate. We set as
\begin{equation}
r_M=0.06w+0.34\quad \text{and}\quad r_F=0.03w+0.17
\end{equation}
so that $r_M$ and $r_F$ slightly increases with $w$. Note that this agrees with (9) when $w=1$.

\begin{figure}[h!]
	\begin{subfigure}{0.47\textwidth}
		\includegraphics[width=\textwidth]{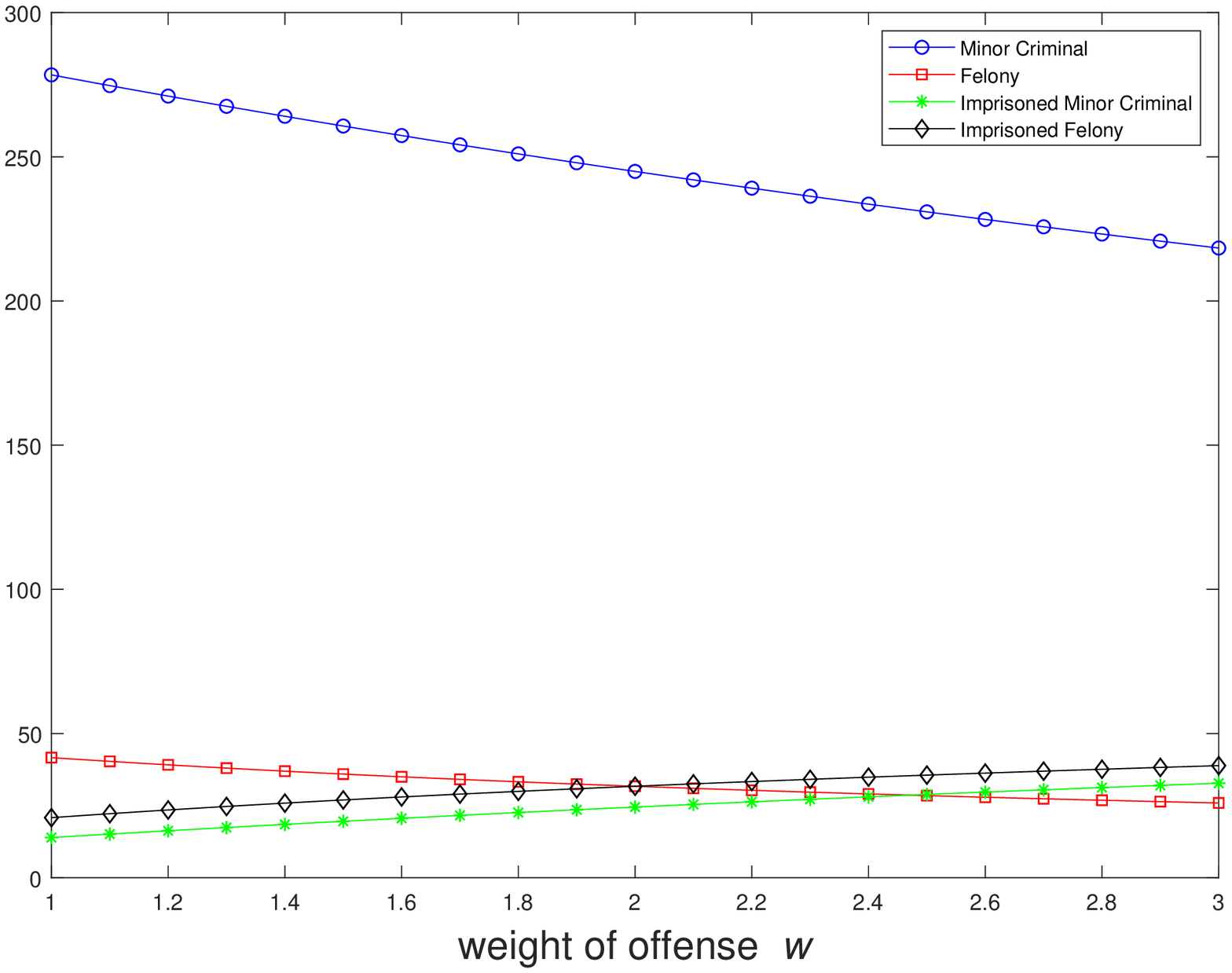}
		\caption{$\beta=0.0002$}
		\centering
	\end{subfigure}
%	\hfill
	\begin{subfigure}{0.47\textwidth}
		\includegraphics[width=\textwidth]{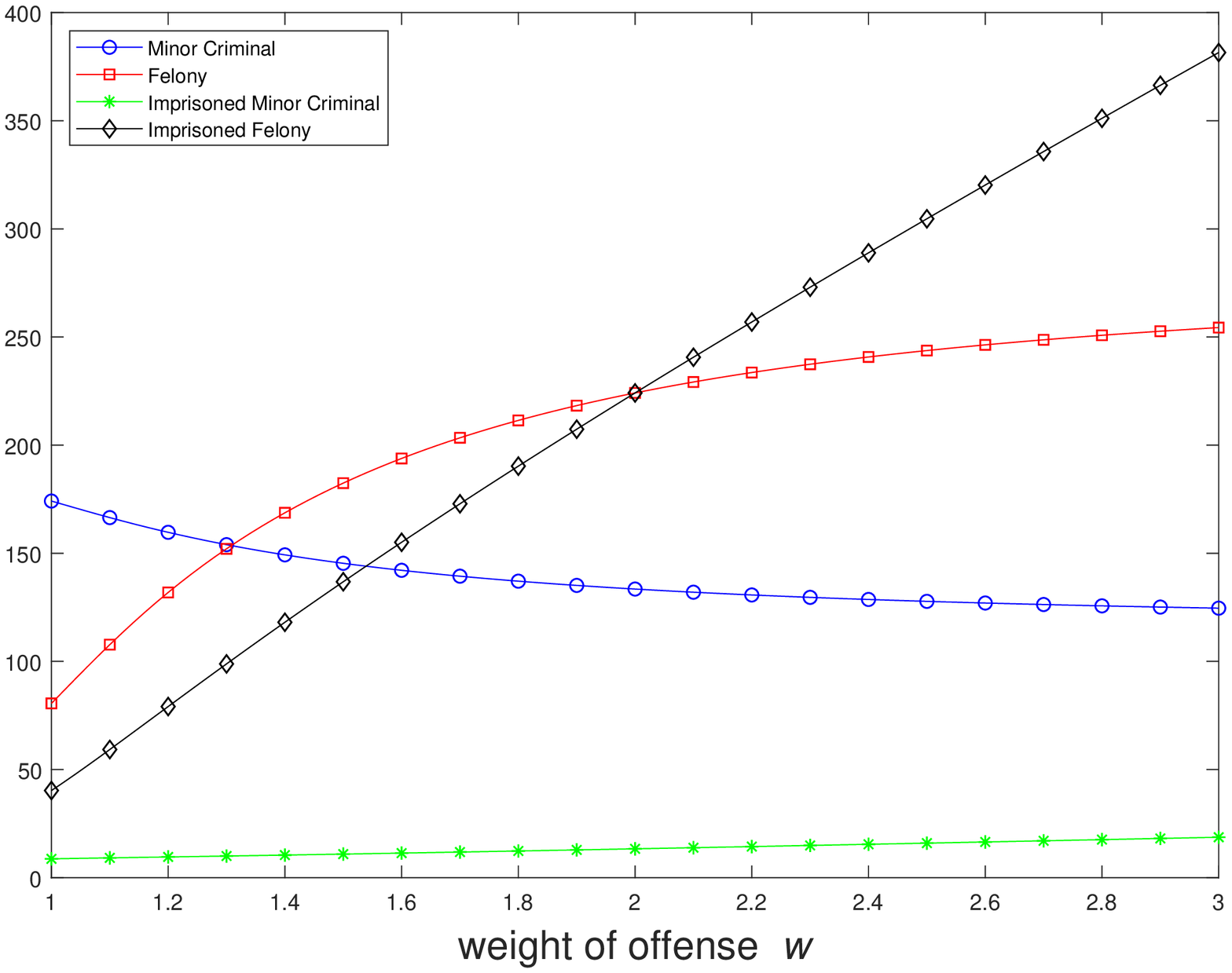}
		\caption{$\beta=0.001$}
		\centering
	\end{subfigure}
\vspace*{8pt}
	\caption{Equilibrium distribution according to the weight of offense $w$ (a) at $\beta=0.0002$ and (b) at $\beta=0.001$.}
\end{figure}

Figure 4 shows how the weight of offense contributes to the criminal distribution in two cases: (a) with $\beta=0.0002$ and (b) $\beta=0.001$. It is no surprising that the number of inmates increases with $w$ in both cases, since a higher weight of offense means a longer detention. More noteworthy differentiation between (a) and (b) is the change in $F$, the number of major criminals in society, according to $w$. When the transmission contact rate is as low as $\beta=0.0002$, a higher weight of offense reduces the number of criminals. On the contrary, when $\beta=0.001$, assigning more weight of offense leads to increase of major crimes in society. Hence a higher weight of offense has a positive reform effect only when the frequent contact between prisoners is effectively prohibited.

Changes in the security measures are likely to have a greater impact on crime \cite{lee2016conclusions}. Let us investigate how the allocation of the police resource to control activity of major/minor crime affects the distribution of criminals. Let $c_{T}$ be the total budget for security. Also let $c_{M}$ and $c_{F}$ be the budget for control of minor crime and major crime, respectively. Note $c_{T}=c_{M}+c_{F}$. We assume that the arrest rate is proportional to the budget used to control the crime. Then we can set $a_{M}=e_{M}c_{M}$ and $a_{F}=e_{F}c_{F}$ where $e_{M}$ and $e_{F}$ are the police efficiency for minor and major crime, respectively. In the example, $e_{M}=0.2$ and $e_{F}=0.04$ are used. Figure 5 shows how the budget ratio $c_{F}/c_{T}$ affects the number of major criminals. The minimum of $F$ is achieved at around $c_{F}/c_{T}\approx 0.4$. Spending more portion of the budget for the major crime control brings negligence on minor crimes, which eventually leads to excessive occurrence of major crimes due to the broken windows effect and the crime school effect.

\begin{figure}[h!]
	\includegraphics[width=0.7\textwidth]{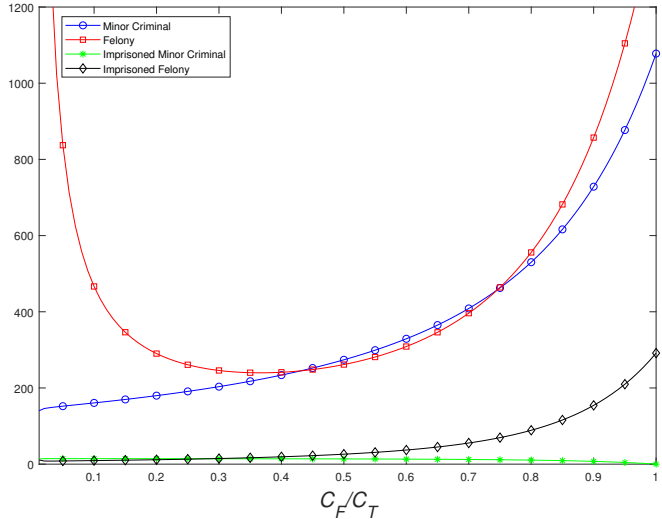}
	\vspace*{8pt}
	\centering
	\caption{Equilibrium distribution according to the allocation of police resource. $e_M = 0.2, e_F = 0.04$}
\end{figure}

\section{Discussion}
We here present the mathematical models for crime dynamics that mainly focus on transition from minor to major criminals occurring in and out of prisons. It is confirmed that both the broken windows effect and the crime-school effect greatly change the criminal distribution. While utilizing the broken windows effect is a preventive measure, improving conditions in correctional facilities can provide a more direct and efficient measure against crimes. The presented work showed that suppressing interactions between overcrowded inmates in prisons is crucial in controlling crimes in society. If not keeping the in-prison contact rate at a low level, extension of the period of imprisonment only results in rapid increase in major crimes.  

The model also shows the importance of an balanced resource allocation between control activity devoted to serious crimes and that devoted to minor crimes. The analysis confirms that, due to the broken windows effect and the crime-school effect, targeting only major crimes can be very inefficient and even bring an opposite result that increases major criminals.  While the results in this work are not predictions, we hope that they can provide useful insights into crime dynamics and possibly suggest effective policies towards crime abatement.

\bigskip

{\bf \large Acknowledgements} \\
This work was supported by the Ministry of Education of the Republic of Korea and the National Research Foundation of Korea (NRF-2017R1D1A1B04032921). The funder had no role in study design, data collection and analysis, decision to publish, or preparation of the manuscript.

\bigskip

\bibliographystyle{amsplain}

\providecommand{\bysame}{\leavevmode\hbox to3em{\hrulefill}\thinspace}
\providecommand{\MR}{\relax\ifhmode\unskip\space\fi MR }
% \MRhref is called by the amsart/book/proc definition of \MR.
\providecommand{\MRhref}[2]{%
  \href{http://www.ams.org/mathscinet-getitem?mr=#1}{#2}
}
\providecommand{\href}[2]{#2}

\end{document}